\RequirePackage{fix-cm}
\documentclass[smallextended]{svjour3}       
\smartqed  
\usepackage{graphicx}
\usepackage{amsmath}
\usepackage{amssymb}
\begin{document}

\title{Zero Variance Markov Chain Monte Carlo\\ for Bayesian Estimators
}


\author{Antonietta Mira \and Reza Solgi \and Daniele Imparato}


\institute{A. Mira \at
              Swiss Finance Institute, University of Lugano, via Buffi 13, CH-6904 Lugano, Switzerland.  \\
              \email{antonietta.mira@usi.ch}           
           \and
           R. Solgi \at
              Swiss Finance Institute, University of Lugano, via Buffi 13, CH-6904 Lugano, Switzerland. \\
\email{reza.solgi@usi.ch}           %
\and D. Imparato \at
Department of Economics, University of Insubria, via Monte Generoso 71, 21100 Varese, Italy. \\
\email{daniele.imparato@uninsubria.it}
}

\date{}

\maketitle

\begin{abstract}
Interest is in evaluating, by Markov chain Monte Carlo (MCMC) simulation,
the expected value of a function with respect
to a, possibly unnormalized, probability distribution.
A general purpose variance reduction technique for the MCMC estimator, based on the zero-variance principle introduced in the physics literature, is proposed.
Conditions for asymptotic unbiasedness of the zero-variance estimator are derived. A central limit theorem is also proved under regularity conditions.
The potential of the idea is illustrated with real
applications to  probit, logit and GARCH  Bayesian models.
For all these models, a central limit theorem and unbiasedness for the zero-variance estimator are proved (see the supplementary material available on-line).
\end{abstract}

\section{General idea}\label{intro}
\noindent
The expected value of
a function $f$ with respect to a, possibly unnormalized, probability
distribution $\pi$,

$\mu_{f}=\int f(\mathbf x)\pi(\mathbf  x)d\mathbf  x/\int\pi(\mathbf x)d\mathbf x$
is to be evaluated.
Markov chain Monte Carlo (MCMC) methods
estimate integrals using a large but finite set of points,
$\mathbf x^{i},i=1,\cdots,N$, collected along the sample path of an ergodic
Markov chain having $\pi$ (normalized) as its unique stationary
and limiting distribution $\hat{\mu}_{f}=\sum_{i=1}^{N}f(\mathbf x^{i})/N$.

In this paper a general method is suggested to reduce the MCMC error by replacing $f$ with a different
function, $\tilde{f}$, obtained by properly re-normalizing $f$.
The function $\tilde{f}$ is constructed so that its expectation,
under $\pi$, equals $\mu_{f}$, but its variance with respect to
$\pi$ is much smaller.
To this aim, a standard variance reduction technique
introduced for Monte Carlo (MC) simulation, known as control variates \cite{RipleySimulation87}, is exploited.

In the rest of this section we briefly explain the zero-variance (ZV) principle introduced in \cite{AssarafCaffarel99,AssarafCaffarel2003}:
an almost automatic method to construct control variates for MC simulation, in which an operator, $H$, acting as a map from functions to functions, and a trial function, $\psi$, are introduced.

In quantum mechanics, a commonly used operator $H$ is the so-called Hamiltonian,
which represents the total energy of the system, that is, the sum of the kynetic energy and the potential energy, where the kinetic energy is typically defined as a second-order differential operator.
Such operator is Hermitian (that is, self-adjoint) if it acts on the restricted class of infinitely differentiable functions with compact support.
If the trial function $\psi$ belongs to this class, and if
\begin{equation}H\sqrt\pi = 0\label{h}\end{equation}
the re-normalized function defined as
\begin{equation}
\tilde{f}(\mathbf x)=f(\mathbf x)+\frac{H\psi}{\sqrt{\pi(\mathbf x)}}
\label{fham} \end{equation}
satisfies
$\mu_{f}=\mu_{\tilde{f}}$:
thus both  $f$ and $\tilde{f}$ can be used to estimate
the desired quantity via Monte Carlo or MCMC simulation.
However, for general $\psi$ the condition $\mu_{f}=\mu_{\tilde{f}}$ may not hold anymore and ad-hoc assumptions on the target $\pi$ are necessary: this issue will be further discussed in Section \ref{sec_unb}.

Inspired by this physical setting, as a general framework $H$ is supposed to be a Hermitian operator (self-adjoint and real in all practical applications) satisfying
\eqref{h}, and the re-normalized function is defined as in \eqref{fham}: depending on the specific
choices of $H$ and $\psi$, the condition $\mu_{f}=\mu_{\tilde{f}}$ has to be carefully verified.

Only a few operators will be considered in the paper, the key one being the Hamiltonian
differential operator. An other
important example discussed below is the Markov operator $H$ acting as
$H\psi(\mathbf{x})=\int K(\mathbf{x},\mathbf{y})\psi(\mathbf{y})d\mathbf{y}$,
where $K(\mathbf{x},\mathbf{y})$ needs to be symmetric.
The re-normalized function, in this case,
 becomes
\begin{equation}
\tilde{f}(\mathbf x)=f(\mathbf x)+\frac{\int K(\mathbf x,\mathbf y)\psi(\mathbf y)d\mathbf y}{\sqrt{\pi(\mathbf x)}}.
\label{f} \end{equation}
and the condition $\mu_{f}=\mu_{\tilde{f}}\label{equal}$ holds as a simple consequence of \eqref{h}.

Regardless of the specific choice of the operator and of the trial function,
the optimal pair $(H,\psi),$ i.e. the one that leads to zero variance,
can be obtained by imposing that $\tilde{f}$ is constant and equal
to its average, $\tilde{f}=\mu_{f}$, which is equivalent to require that
$\sigma^2(\tilde{f})=0$, where $\sigma^2(\cdot)$ denotes the variance operator with respect to the target $\pi$.
The latter, together with (\ref{fham}), leads to the fundamental equation:
\begin{equation}
H\psi=-\sqrt{\pi(\mathbf x)}[f(\mathbf x)-\mu_{f}].\label{fun}\end{equation}
 In most practical applications equation (\ref{fun}) cannot be solved
exactly,
still, we propose to find an approximate solution in the following
way. First choose a Hermitian operator $H$ verifying (\ref{h}).
Second, parametrize
$\psi$ and derive the optimal parameters by minimizing $\sigma^2(\tilde{f})$.
The optimal parameters are then estimated using a first short MCMC simulation.
Finally, a much longer MCMC simulation is performed using $\hat{\mu}_{\tilde{f}}$
instead of $\hat{\mu}_{f}$ as the estimator. This final estimator will be called
Zero Variance (ZV) estimator through the paper.

Other research lines aim at reducing the asymptotic variance of MCMC
estimators by modifying the transition kernel of the Markov chain. These modifications have been achieved in many different ways, for example
by trying to induce negative correlation along the chain path (\cite{BaroneFrigessi89,GreenHan92,CraiuMeng05,So06,CraiuLemieux07}); by trying to avoid random walk behavior via
successive over-relaxation (\cite{Adler1981,Neal1995,Baronealii2001});
by hybrid Monte Carlo (\cite{Duanealii1987,Neal1994,Breweralii1996,Fortalii2003,Ishwaran1999}); by exploiting non reversible Markov chains (\cite{Diaconisalii2000,MiraGeyer00}), by delaying rejection in Metropolis-Hastings type algorithms (\cite{TierneyMiraAdaptive99,GreenMiraRJ01}),
by data augmentation (\cite{Vandykalii2001,GreenMiraRJ01}) and
auxiliary variables (\cite{SwendsenWang1987,Higdon1998,Miraalii2001,TierneyMira2002}). 
Up to our knowledge, the only other research line that uses control variates in MCMC estimation
follows
the PhD thesis
by \cite{Henderson97} and has its most recent developement in \cite{Dellaportas2010}.  In \cite{HendersonGlynn2002} it is observed that, for any real-valued function $g$ defined on the state space of a Markov chain $\{X^n\}$, the
one-step conditional expectation
$U(\mathbf{x}) := g(\mathbf{x}) - \mathbb E [g(X^{n+1})|X^n = \mathbf{x} ]$ has zero mean with respect to the stationary distribution of the chain and can thus be used as control variate.
The Authors also note that the best choice for the function $g$
is the solution of the associated Poisson equation which can rarely be
obtained analytically but can be approximated in specific settings.
In \cite{Dellaportas2010}, the use of this type of control variates is further explored in
the setting of reversible Markov chains were a closed form expression for $U$ is often available.

In \cite{AssarafCaffarel99,AssarafCaffarel2003}
unbiasedness and existence of a central limit theorem (CLT) for the ZV estimator are not discussed, neither in \cite{Leisen2010}, where this estimator is applied to a toy example. The main contributions of this paper are, on the one hand, to derive the rigorous conditions for unbiasedness and CLT for the ZV estimators in MCMC simulation. On the other hand, we apply the ZV principle to some widely used models (probit, logit, and GARCH) and demonstrate that, under very mild restrictions, the necessary conditions for unbiasedness and CLT are verified.

\section{Choice of $H$}
\label{hh}In this section guidelines to choose the operator $H$,
both for discrete and continuous
settings, are given.
In a discrete state space, denote with $P(\mathbf x,\mathbf y)$ a transition matrix reversible with respect
to $\pi$ (a Markov chain will be identified with the corresponding transition
matrix or kernel).
We restrict our attention in this section
to operators $H$ acting as
$Hf:=\sum_y K(\mathbf{x},\mathbf{y})f(\mathbf{y})$. The following choice
\begin{equation}
K(\mathbf x,\mathbf y)=\sqrt{\frac{\pi(\mathbf x)}{\pi(\mathbf y)}}[P(\mathbf x,\mathbf y)-\delta(\mathbf x-\mathbf y)]\label{dellap}\end{equation}
satisfies condition (\ref{h}), where
$\delta(\mathbf x-\mathbf y)$ is the Dirac delta
function: $\delta(\mathbf x-\mathbf y)=1$ if $\mathbf x=\mathbf y$ and zero otherwise. It should
be noted that the reversibility condition imposed on the Markov chain is essential
in order to have a symmetric operator $K(\mathbf x,\mathbf y)$, as required.

With this choice of $H$, letting $\tilde{\psi}=\psi/\sqrt{\pi}$, equation
(\ref{f}) becomes: \[
\tilde{f}(\mathbf x)=f(\mathbf x)-\sum_{\mathbf y} P(\mathbf x,\mathbf y)[\tilde{\psi}(\mathbf x)-\tilde{\psi}(\mathbf y)].\]
The same $H$ can also be applied in continuous settings. In this case,
$P$ is the kernel of the Markov chain and equation (\ref{dellap}) can be trivially extended. This choice of $H$ is exploited in
\cite{Dellaportas2010}, where the following fundamental
equation is found for the optimal $\tilde\psi$:
$\mathbb E[\tilde\psi(\mathbf x_1)|\mathbf x_0=\mathbf x]-\tilde\psi(\mathbf x)=\mu_f-f(\mathbf x)$.
It is easy to prove that this equation coincides with our fundamental equation (\ref{fun}),
with the choice of $H$ given in (\ref{dellap}). The Authors observe that the optimal trial function is given by
\begin{equation}
\tilde\psi(\mathbf x)=\sum_{n=0}^\infty[\mathbb E[f(\mathbf x_n)|\mathbf x_0=\mathbf x]-\mu_f] \label{psidellaportas},
\end{equation}
that is, $\tilde\psi$ is the solution to the Poisson equation for $f(\mathbf x)$. However, an explicit solution cannot
be obtained in general.

Another operator is proposed in \cite{AssarafCaffarel99}: \label{Hnuovo} if
$\mathbf x\in\mathbb R^{d}$ consider the Schr\"odinger-type Hamiltonian
operator: \begin{equation}
H f=-\frac{1}{2}\sum_{i=1}^{d}\frac{\partial^{2}}{\partial
x_{i}^{2}}f+V(\mathbf x)f\label{cont1},\end{equation}
 where $V(\mathbf x)$ is constructed to fulfill equation (\ref{h}):
$ V=\frac1{2\sqrt{\pi}}\Delta\sqrt{\pi}$ and $\Delta$ denotes the Laplacian operator of second order derivatives. In this setting,
we obtain the general expression for $\tilde f$ reported in \eqref{fham}, where now $H$ is the Schr\"odinger-type Hamiltonian.
 These are the operator and the re-normalized function that will be considered throughout this paper.
Although it can only be applied to continuous state spaces, this Schr\"odinger-type operator
shows several advantages with respect to the operator (\ref{dellap}). First of all, in order to use (\ref{dellap})
the conditional expectation appearing in (\ref{psidellaportas}) has to be available in closed form. Secondly, definition
(\ref{cont1}) does not require reversibility of the Markov chain. Moreover, this definition is independent
of the kernel $P(\mathbf x,\mathbf y)$ and, therefore, also of the type of MCMC algorithm that is used in the simulation.
Note that, for calculating $\tilde{f}$ both with the operator (\ref{cont1}) and (\ref{dellap}),
the normalizing constant of $\pi$ is not needed.


\section{Choice of $\psi$}

\label{choicefi}

The optimal choice of $\psi$ is the exact solution of the
fundamental equation (\ref{fun}). In real applications, typically,
only approximate solutions, obtained by minimizing
$\sigma^2(\tilde{f})$, are available. In other words, we select a
functional form for $\psi$, parameterized
by some coefficients of a class of polynomials, and optimize those coefficients by
minimizing the fluctuations of the resulting $\tilde{f}.$ The
particular form of $\psi$ is very dependent on the problem at
hand, that is on $\pi$, and on $f$. In the sequel it will be assumed that $\psi=P\sqrt\pi$, where P is a
polynomial.
As one would expect, the higher is the degree of the polynomial, the higher is the number of control
variates introduced and the higher is the variance reduction achieved. It can be easily shown that in a $d$ dimensional space, using polynomials of order $p$, provides ${d+p \choose d}-1$ control variates. However,
some restrictions on the coefficients may occur in order to get an unbiased MCMC estimator. See Example \ref{exa_unb} of Section \ref{sec_unb} at this regard.

\section{Control Variates and optimal coefficients}\label{cv}
In this section, general expressions for the control variates in the ZV method are derived.
Using the Schr\"odinger-type Hamiltonian $H$ as given in (\ref{cont1})
and trial function $\psi(\mathbf x) = P(\mathbf x) \sqrt{\pi(\mathbf x)}$,
the re-normalized function is:
\begin{eqnarray}
   \tilde{f}(\mathbf x) = f(\mathbf x) - \frac{1}{2} \Delta P(\mathbf x)  + \nabla P(\mathbf x) \cdot\mathbf  z, \label{renorm}
\end{eqnarray}
where
$ \mathbf   z =  - \frac{1}{2} \nabla \ln \pi(\mathbf x)$,
$\nabla = \left(\frac{\partial}{\partial x_1}, ..., \frac{\partial}{\partial x_d}\right)$ denotes the gradient and $\Delta = \sum_{i=1}^{d} \frac{\partial^2}{\partial x_i^2}$. Like any other control variate (i.e. zero mean random variables under the distribution of interest), the variable $\mathbf z$ can be monitored to test convergence along the lines suggested by \cite{BrooksGelman1998} and \cite{PhilippeRobert01}, where the same control variate
$ \mathbf z = \nabla \log\pi$ is used.

Hereafter the function $P$ is assumed to be a polynomial.
As a first case, for $P(\mathbf x) = \sum_{j=1}^{d} a_j x_j$ (1st degree polynomial), one gets:
\begin{equation}
   \tilde{f}(\mathbf x) = f(\mathbf x) + \frac{H \psi(\mathbf x)}{\sqrt{\pi(\mathbf x)}} = f(\mathbf x) + \textbf{a}^T \textbf{z}.  \nonumber
\end{equation}
The optimal choice of $\textbf{a}$, that minimizes the variance of $\tilde{f}(x)$, is:
\begin{eqnarray}
     \textbf{a} = - \Sigma_{\textbf{zz}}^{-1} \sigma(\textbf{z},f),  \ \ \ \ \ \ \textmd{where} \ \ \ \ \ \ \ \Sigma_{\textbf{zz}} = \mathbb{E}(z z^T), \ \  \sigma(\textbf{z}, f) = \mathbb{E}(z f). \nonumber
\end{eqnarray}
For a more general approach to the choice of coefficients using control variates, reference should be made to \cite{Nelson1989} and \cite{Loh1994}.
We anticipate that conditions under which the ZV-MCMC estimator obeys a CLT (Section \ref{sec_unb}) guarantee that
the optimal $\mathbf a$ is well defined.
In ZV-MCMC, the optimal $\textbf{a}$ is estimated in a first stage, through a short MCMC simulation\footnote{From a practical point of view there is no need to run two separate chains, one to get the control variates and one to get the final ZV estimator: everything can be done on a single Markov chain which is run once to estimate the optimal coefficients of the control variates and then post-processed to get the ZV estimator.}.
When higher-degree polynomials are considered, a similar formula for the coefficients associated to the
control variates is obtained once
an explicit formula for the control variate vector $\mathbf z$ has been found.
As an example, for quadratic polynomials
     $P(\mathbf x) =\mathbf  a^T \mathbf x + \frac{1}{2} \mathbf x^T B \mathbf x$,
the re-normalized $\tilde f$ is :
\[
   \tilde{f}(\mathbf x) = f(\mathbf x) -\frac{1}{2} \textmd{tr}(B) + (\mathbf a +  B \mathbf x)^T \mathbf z.\]  	
Using second order polynomials yields a vector of control variates of dimension $\frac12 d(d+3)$. Therefore, finding the optimal coefficients requires working with $\Sigma_{zz}$ which is a matrix of dimension of order$d^2$. This  makes the use of second order polynomials computationally expensive
when dealing with high-dimensional sampling spaces, say of the order of decades.
\section{Unbiasedness and central limit theorem}\label{sec_unb}
As remarked in Section \ref{intro}, condition (\ref{h}) may not be sufficient to ensure
unbiasedness of the estimator when the Schr\"odinger operator (\ref{cont1}) is used. In this section general conditionys on the target $\pi$ are
provided that guarantee that the ZV-MCMC estimator is
(asymptotically) unbiased for the class of trial functions discussed. Details can be found in the on-line supplementary material, Appendix D.
\begin{proposition}
Let $\pi$ be a $d$-dimensional density
on a bounded open set $\Omega$ with regular boundary $\partial \Omega$,
whose first and second derivatives are continuous. Then, if $\psi=P\sqrt\pi$,
a sufficient condition for unbiasedness of the ZV-MCMC
estimator is
$\pi(\mathbf x)\frac{\partial P (\mathbf x)}{\partial x_j}=0$, for all $\mathbf x\in\partial\Omega$, $j=1,\ldots, d$.
\end{proposition}
The previous proposition is a consequence of multidimensional integration by parts, from which one gets the equality
\begin{equation}
\mathbb E_\pi\left[\frac{H\psi}{\sqrt\pi} \right]=\frac12
\displaystyle\int_{\partial\Omega}[\psi\nabla\sqrt\pi-\sqrt\pi\nabla\psi]\cdot
\mathbf n d\sigma, \label{cond_unb}\end{equation}
where $\mathbf n$ denotes the
versor orthogonal to $\partial \Omega$.

When $\pi$ has unbounded support, integration by
parts cannot be used directly. In this case, we can formulate the following result.
\begin{proposition}
Let $\pi$ be a $d$-dimensional density
with unbounded support $\Omega$, whose first and second derivatives are continuous, and let  $(B_r)_r$ be
a sequence of bounded
subsets,
so that
$B_r\nearrow\Omega$. Then, a sufficient condition for unbiasedness of the ZV-MCMC estimator is
\[\lim_{r\rightarrow+\infty} \displaystyle\int_{\partial
B_r}\pi\nabla P\cdot \mathbf n d\sigma=0.\]
\end{proposition}
In the univariate case, if
$\Omega$ is some interval of the real line, that is,
$\Omega=(l,u)$, where $u,l\in\mathbb R
\cup\pm\infty$, it is sufficient that \begin{equation}\frac{d
P(x)}{dx}\bigg|_{x=l}\pi(l)=\frac{d P(x)}{dx}\bigg|_{x=u}\pi(u),
\label{unb}\end{equation} which is true, for example, if
$\frac{dP}{dx}\pi$ annihilates at the border of the support.

In the seminal paper by \cite{AssarafCaffarel99} unbiasedness conditions are not clearly explored since, typically, the target distribution the physicists are interested in, annihilate at the border of the domain
with an exponential rate.
The following example shows how crucial the choice of trial
functions is, in order to have an unbiased estimator, even in
trivial models.
\begin{example}
Let $f(x)=x$ and $\pi$ be exponential: $\pi(x)=\lambda
e^{-\lambda x}\mathbb I_{\{x>0\}}$. If $P(x)$ is a first order
polynomial, (\ref{unb}) does not hold and this
choice does not allow for a ZV-MCMC estimator, since the control
variate $z=-\frac12 \frac{d}{dx} \ln\pi(\mathbf x)$ is constant and
$\sigma(x,z)=0$. However, to satisfy equation
(\ref{unb}) it is sufficient to consider second order polynomials.
Indeed, if $P(x)=a_0+a_1x+a_2x^2$ equation (\ref{unb}) is satisfied
provided that $a_1=0$ and the minimization of the variance of $\tilde f$ can be carried
out within this special class. The optimal
choice $a_2:=\frac1{2\lambda}$ yields zero variance:
$\sigma^2(\tilde f)\equiv 0$.
\label{exa_unb}\end{example}
\subsection{Central limit theorem}
Conditions for existence of a CLT for $\hat{\mu}_f$ are well known in the literature  (\cite{TierneyMCMC94}).
Using these classical results, from \eqref{renorm} we have that
the ZV-MCMC estimator obeys a CLT provided $f$,
$\Delta P$ and $\nabla P\cdot \mathbf z$ belong to  $L^{2+\delta}(\pi)$ when the Markov chain
run for the simulation is geometrically
ergodic.
In the next corollary, the case of linear and quadratic polynomials $P$ (used in the examples in Section \ref{exa}) is considered.
\begin{corollary}
Let $\psi(\mathbf x)=P(\mathbf x)\sqrt{\pi}$, where $P(\mathbf x)$ is a first or second degree polynomial. Then,
the ZV-MCMC
estimator $\hat{\mu}_{\tilde{f}}$ is a consistent estimator of $\mu_f$ which satisfies the CLT, provided one of the following conditions holds:
    \begin{enumerate}
                      \item[{C1}]:
                 The Markov chain is geometrically ergodic and $f$, $x_i^k z_j \in L^{2+\delta}(\pi)$,
$\forall i,j$, for all
$k\in\{0, \deg P-1\}$ and some $\delta > 0$.
            \item[{C2}]:
                 The Markov chain is uniformly ergodic and $f$, $x_i^kz_j\in L^{2}(\pi)$,
$\forall i,j$ and for all
$k\in\{0,\deg P-1\}$.
        \end{enumerate}
\end{corollary}
In the case of linear $P$, using the definition of control variate,
the statement of the previous corollary can be
reformulated in this simple way:
if  $f\in L^2(\pi)$ and the chain is uniformly ergodic, then
a sufficient condition to get a CLT is
\begin{equation}
   m_j = \mathbb{E}_{\pi}\left[\left(\frac{\partial}{\partial x_j} \ln(\pi(\mathbf x))\right)^2\right]
             <\infty, \ \ \ \forall j.
\nonumber \end{equation}
The quantity $m_j$ is known in the literature as Linnik functional (if considered as a function of the target distribution, $I(\pi)$) since
it was introduced by \cite{Linnik59}.
The quantity $m_j$ is also interpretable as the Fisher
information of a location family in a frequentist setting.
\subsection{Exponential family}
Let $\pi$ belong to a $d$-dimensional exponential family:
$\pi(\mathbf x)\propto \exp(\beta\cdot \mathbf T(\mathbf x)-K_p(\beta))p(\mathbf
x)$,
where
$\beta\in\mathbb R^d$ is the vector of natural parameters.
The following theorem provides a sufficient condition for a CLT for ZV-MCMC estimators when the target belongs to the exponential family and a
uniformly ergodic Markov Chain is considered.
Similar results can be achieved
when the Markov Chain is geometrically ergodic, by considering the $2+\delta$ moment. This statement can be easily
verified by a direct computation.
\begin{theorem}
Let $\pi$ belong to an exponential family, with $p$
such that $\frac{\partial \log p}{\partial x_j}\in L^2(\pi)$, $\forall i,k$.
Then, the Linnik functional of $\pi$ is finite
if and only if
$\, \frac{\partial T_k}{\partial x_j}\in L^2(\pi)$, $\forall i,k$.
\label{thmexp}\end{theorem}
\begin{example}
The Gamma density $\Gamma(\alpha,\theta)$ can be written as an exponential family on $(0,+\infty)$,
where $p(x) \equiv 1$,
so that hypotheses of Theorem \ref{thmexp} are satisfied. A direct computation shows that the Gamma density $\Gamma(\alpha,\theta)$ has finite Linnik functional for
any $\theta$ and for any $\alpha \in\{1\}\cup (2,+\infty)$. Under these conditions, a CLT holds for the ZV-MCMC estimator.
\end{example}
\section{Examples}\label{exa}
In the sequel standard statistical models are considered.
For these models, the ZV-MCMC estimators are derived
in a Bayesian context; from now on, the target $\pi=\pi(\beta|\mathbf x)$ is the Bayesian
posterior distribution:
therefore, the argument associated with the state of the Markov chain is denoted by $\beta$ instead of $\mathbf x$, which represents, now, the vector of data.
The operator $H$ considered is the Schr\"odinger-type Hamiltonian defined in (\ref{cont1}), and
$\psi=P\sqrt\pi$, where P is a polynomial.

Numerical simulations are provided, that confirm the effectiveness of variance reduction achieved, by
minimizing the variance of $\tilde f$ within the class of trial functions considered.
Moreover, conditions for both unbiasedness and CLT for $\tilde f $ are verified for all the examples.
For the mathematical derivation of the zero-variance estimator and
the proofs of unbiasedness and CLT for the models considered, we refer the reader to the appendices of the on-line supplementary material (Appendices A, B and C).

\subsection{Probit Model}\label{probit}
\begin{figure*}
\includegraphics[width=12cm]{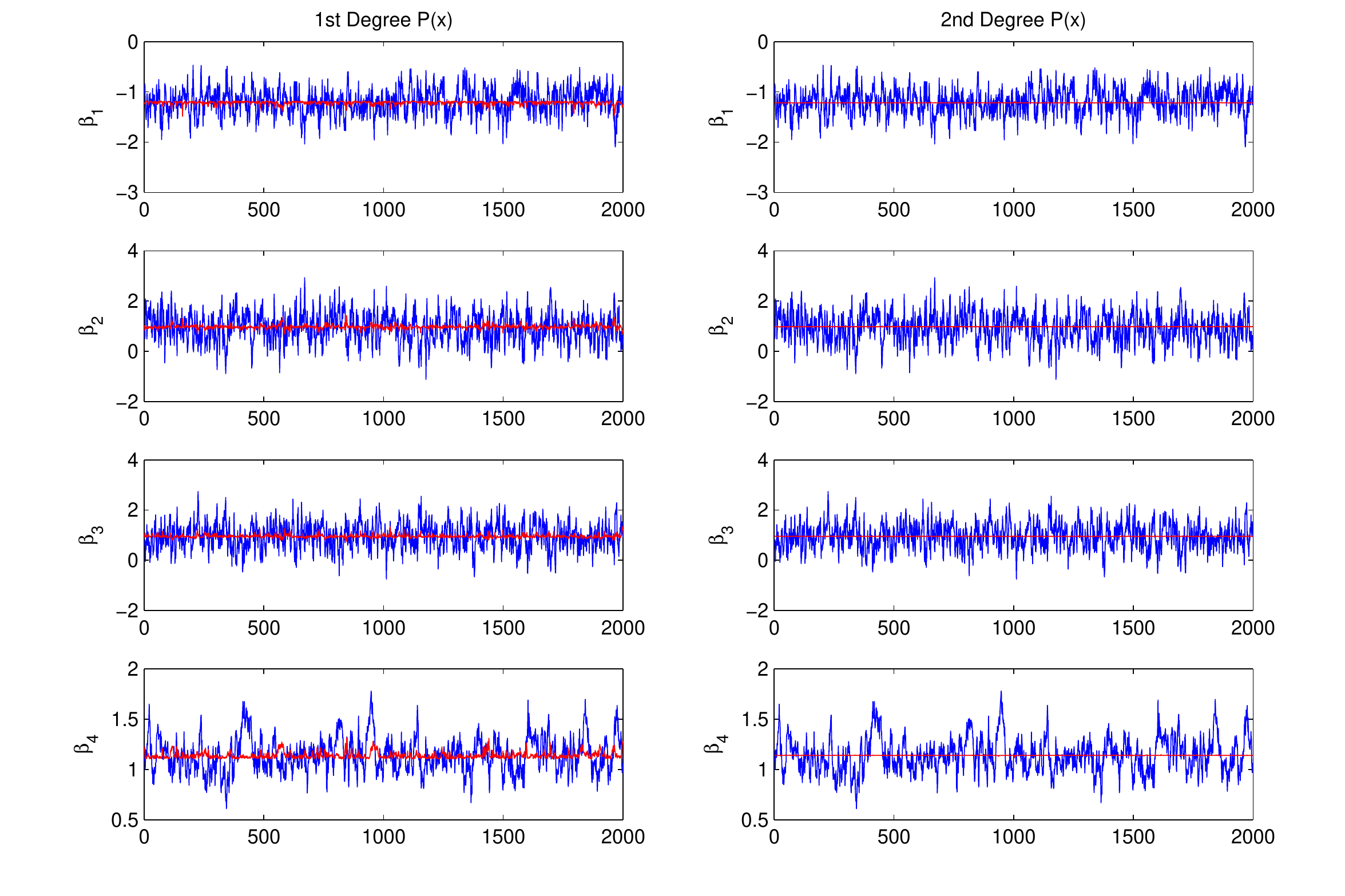}
\caption{Ordinary MCMC (blue) and ZV-MCMC (red) for probit model: rows are
parameters, columns are degree polynomials.}
\label{fig:probit1}
\end{figure*}

To demonstrate the effectiveness of ZV for probit models, a simple example is presented.
The bank dataset from \cite{Flury88} contains the measurements of four variables on 200 Swiss banknotes (100 genuine and 100 counterfeit).
The four measured variables $x_i$ ($i=1, 2, 3, 4$), are the length of the bill, the width of the left and the right edge, and the bottom margin width. These variables are used in a probit model as the regressors, and the type of the banknote $y_i$, is the response variable (0 for genuine and 1 for counterfeit).
Using flat priors, the Bayesian estimator of each parameter, $\beta_k$, under squared error loss function,
is the expected value of
$f_k(\beta) =\beta_k$  under $\pi$ ($k=1, 2, \cdots, d$). The Bayesian analysis of this problem is discussed in \cite{BayesianCore}.
In order to find the optimal
vector of parameters $a_{k}$ of the trial functions,
a short Gibbs sampler, following  (\cite{AlbertChib03}), (of length 2000, after 1000 burn in steps) is run, and the optimal coefficients are estimated:
 $\hat{\textbf{a}}_k = - \hat{\Sigma}_{\textbf{zz}}^{-1} \hat{\sigma}(\textbf{z},\beta_k)$.
Finally another MCMC simulation of length 2000 is run (and using the estimated optimal values obtained in the previous step), along which $\widetilde{f}_k(\beta)$, for $k=1,\ldots, 4$ is averaged. We have repeated this experiment 100 times. The MCMC traces of the ordinary MCMC and the ZV-MCMC in one of these MOnte Carlo experiments have been depicted in the left plot of Fig. \ref{fig:probit1}. The blue curves are the traces of $f_k$ (ordinary MCMC), and the red ones are the traces of $\widetilde{f}_k$ (ZV-MCMC). It is clear from the figure that the variances of the estimator have  substantially decreased.
Indeed for the linear trial functions, the ratios of the Monte Carlo estimates of the asymptotic variances of the two estimators (ordinary MCMC and ZV-MCMC) are between 25 and 100.
Even better performance can be achieved using second degree polynomials to define the trial function. In the right column of Fig. \ref{fig:probit1} the traces of ZV-MCMC with second order $P(x)$ are reported along with the traces of the ordinary MCMC. As it can be seen from the figure, the variances of the ZV estimators are negligible: the ratio of the Monte Carlo estimates of the asymptotic variances of the two estimators are between $18,000$ and $90,000$. In this example (with the simulation length and burn-in reported above) the CPU time of  
ZV-MCMC is almost $3$ times larger than the one of ordinary MCMC.\\

\begin{figure*}
\includegraphics[width=12cm]{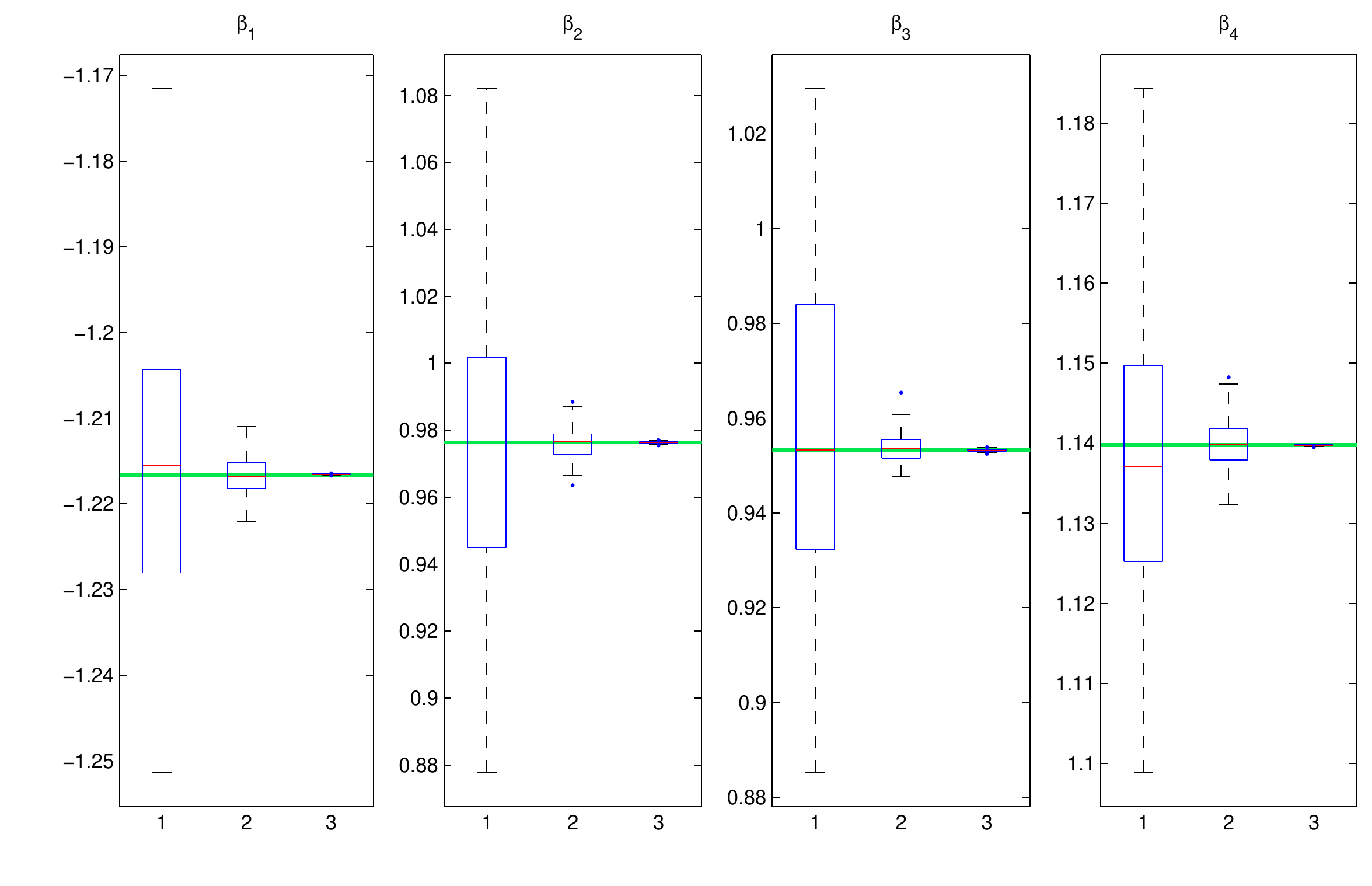}
\caption{Boxplots of ordinary MCMC estimates (1) and ZV-MCMC estimates (2 and 3) for the probit model, along with the $95\%$ confidence region obtained by an ordinary MCMC of length $10^8$ (green regions).}
\label{fig:probitBoxplots}
\end{figure*}

In order to study the unbiasedness of the ZV-estimators empirically, we have run a very long MCMC (of length $10^8$) and obtained a very narrow $95\%$ confidence region for each parameter. In Fig. \ref{fig:probitBoxplots} we have depicted the box-plot of the ordinary MCMC (first box-plot), and the ZV-estimators (second and third box-plot) along with these $95\%$ confidence regions (the green regions). As it can be seen, the ZV-estimators are concentrated in the $95\%$ confidence regions obtained from the very long chain.

\subsection{Logit Model:}
A logit model is fitted to the same dataset of Swiss banknotes previously introduced.
Flat priors are used and, as before, the Bayesian estimator of each parameter, $\beta_k$, (again under squared error loss functions) is the expected value of $\beta_k$  under $\pi$ ($k=1, 2, \cdots, d$).
Similar to the probit example, in the first stage a MCMC
simulation is run, and the optimal parameters of $P(\beta)$ are
estimated.
Then, in the second stage, an independent simulation is performed, and $\tilde{f}_k$ is averaged, using the optimal trial
function estimated in the first stage
(the same simulation length and burn-in, as in the probit example, have been used).
For linear polynomial, the ratio
of the Monte Carlo estimates of the asymptotic variances of the two
estimators (ordinary MCMC and ZV-MCMC) are
between 15 and 50. Using quadratic polynomials, these ratios are between $15,000$ and $20,000$. In this example the CPU time of the ZV-MCMC is almost $3$ times higher than that of ordinary MCMC.\\



\begin{figure*}
\includegraphics[width=12cm]{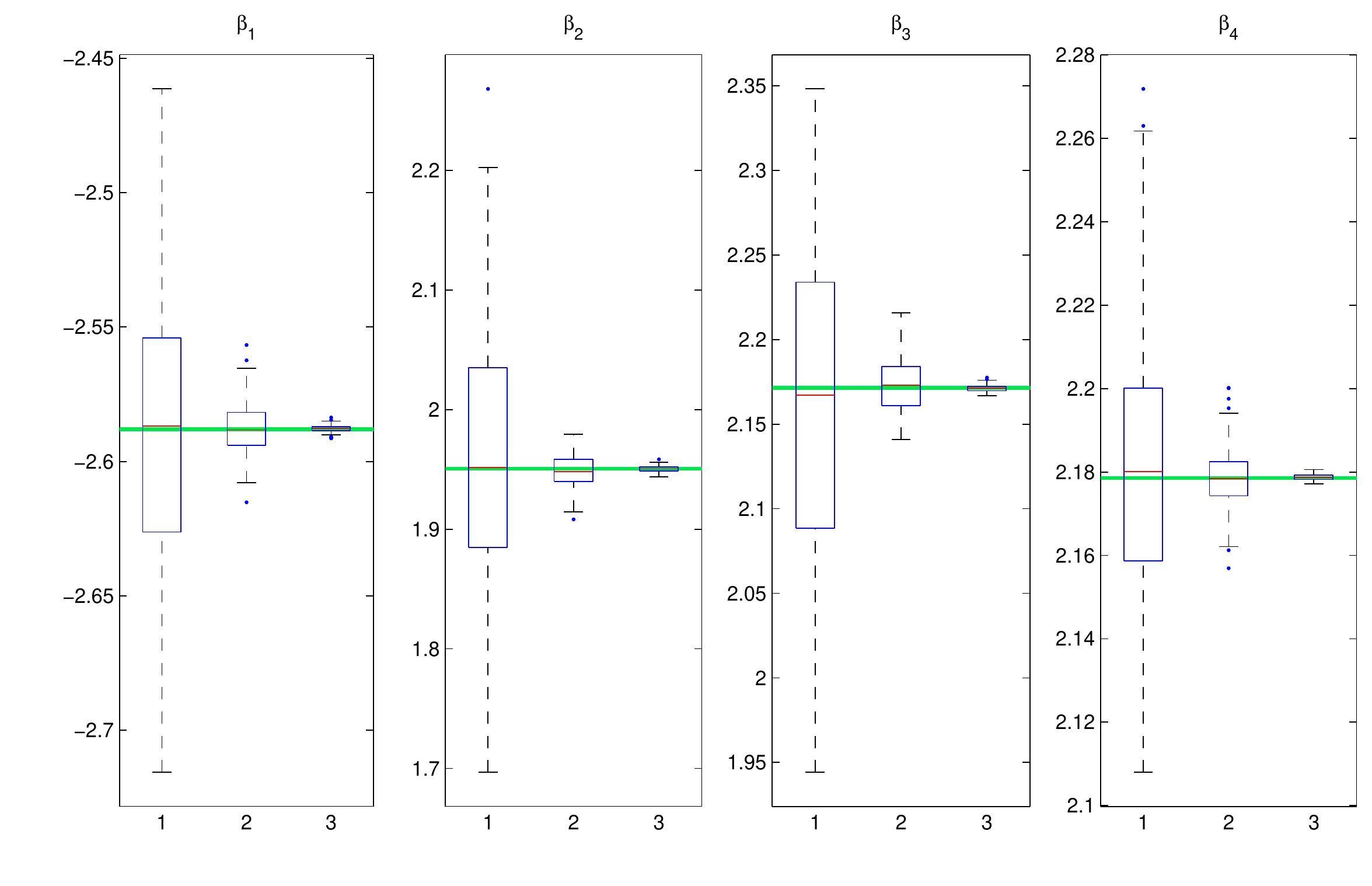}
\caption{Box-plots of ordinary MCMC estimates (1) and ZV-MCMC estimates (2 and 3) for the logit model, along with the $95\%$ confidence region obtained by an ordinary MCMC of length $10^8$ (green regions).}
\label{fig:logitBoxplots}
\end{figure*}

We have run a very long MCMC (of length $10^8$) and obtained a very narrow $95\%$ confidence region for each parameter. In Fig. \ref{fig:logitBoxplots} we have depicted the box-plot of the ordinary MCMC (first box-plot), and the ZV-estimators (second and third box-plot) along with these $95\%$ confidence regions (the green regions).
Again, as it can be seen, the ZV-estimators are concentrated in the $95\%$ confidence regions obtained from the very long Markov chain.

\subsection{GARCH Model}
Generalized autoregressive conditional heteroskedasticity (GARCH) models (\cite{Bollerslev86}) have become one of the most important building blocks of models in financial econometrics,
where they are widely used to model returns. Here it is shown how the ZV-MCMC principle can be exploited to estimate
the parameters of a univariate GARCH model
applied to daily returns of exchange rates in a Bayesian setting.
Let $S(t)$ be the exchange rate at time $t$.
The daily returns are defined as
   $ r(t) := [S(t) - S(t-1)]/S(t-1) \approx \ln\left(S(t)/S(t-1)\right)$.
In a Normal-GARCH model, we assume the returns are conditionally Normally distributed,
   $ r(t) | \mathcal{F}_t \sim \mathcal{N}(0, h_t)$, where
$    h_t = \omega_1 + \omega_3 h_{t-1} + \omega_2 r_{t-1}^2$,
and $\omega_1>0$, $\omega_2 \geq 0$, and $\omega_3 \geq 0$ are the parameters of the model. The aim is to estimate the expected value of $\omega_j$ under the posterior $\pi$, using
independent truncated normal priors.
As an example,  a Normal-GARCH(1, 1) is fitted to the daily
returns of the Deutsche Mark vs British Pound exchange
rates from
January 1985, to December 1987.
In the first stage a short MCMC simulation (\cite{Ardia2008}) is used to estimate the optimal parameters of the trial function (2000 sweeps after 1000 burn-in). In the second stage an
independent simulation is run (with length 10000) and $\tilde{f}_k(\omega)$
is averaged in order to efficiently estimate the posterior mean of each parameter. We compare this ZV-MCMC with an ordinary MCMC of length 10000 (after 1000 burn-in).
First, second and third degree polynomials in the trial function are used. In order to study the effectiveness of ZV-MCMC, we have run these simulations (ordinary MCMC and ZV-MCMC) 100 times.
As it can be seen in Table \ref{table1},
where a 95\% confidence interval for the variance reductions are reported, the ZV strategy reduces the variance of the estimators up to ten thousand times. In this example (with the simulation and burn-in lengths reported above) the CPU time of the ZV-MCMC is almost $20\%$ higher than the CPU time of ordinary MCMC.\\


\begin{figure*}
\includegraphics[width=12cm]{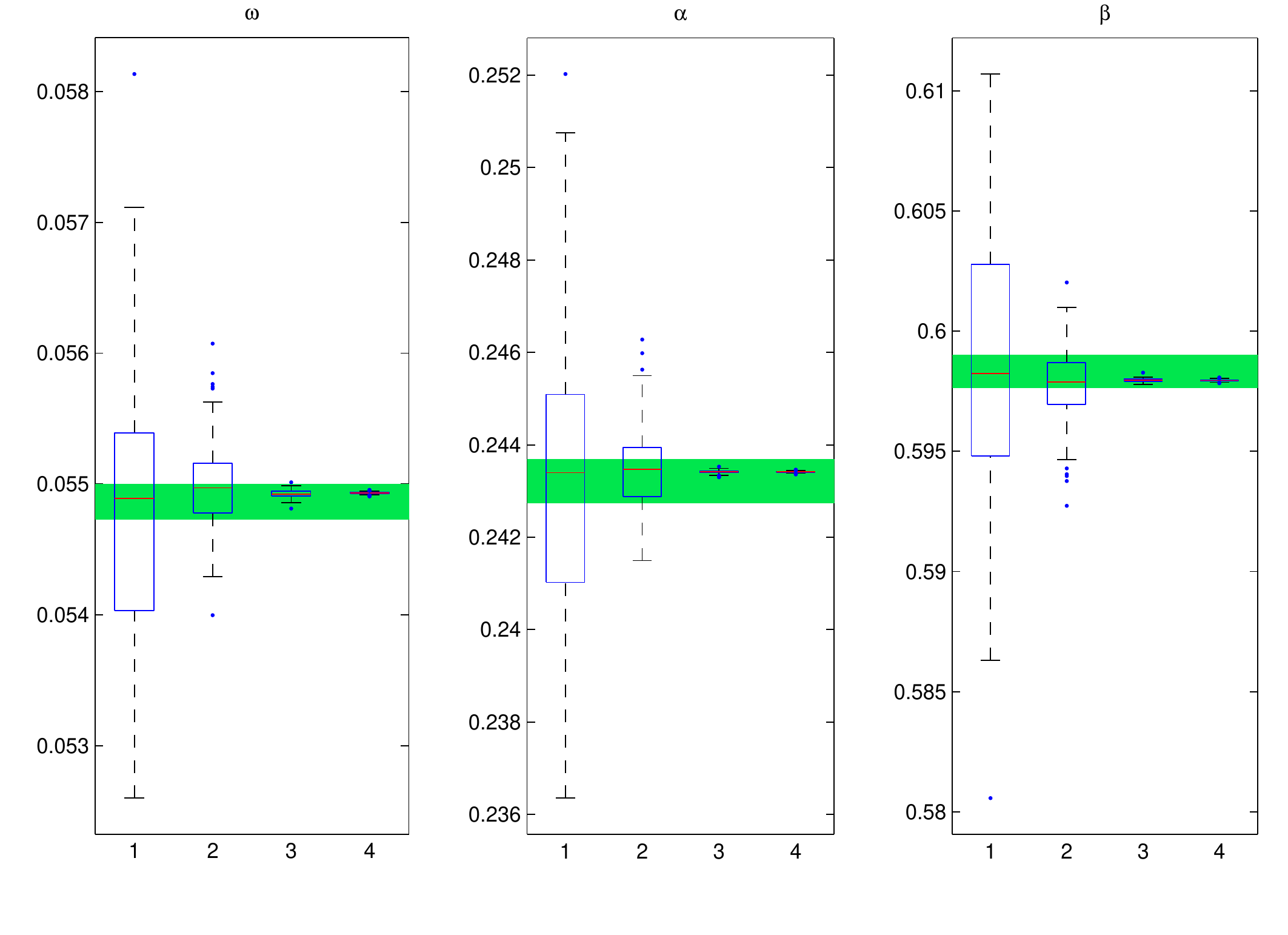}
\caption{Boxplots of ordinary MCMC estimates (1) and ZV-MCMC estimates (2, 3 and 4) for the GARCH model, along with the $95\%$ confidence region obtained by an ordinary MCMC of length $10^7$ (green regions).}
\label{fig:GARCHBoxplots}
\end{figure*}
In order to study the unbiasedness of the ZV-estimators empirically, we have run a very long MCMC (of length $10^7$) and obtained a narrow $95\%$ confidence region for each parameter. In Fig. \ref{fig:GARCHBoxplots} we have depicted the box-plot of the ordinary MCMC (first box-plot), and the ZV-estimators (second, third and fourth box-plots) along with these $95\%$ confidence regions (the green regions). As it can be seen the ZV-estimators lie in the range obtained by the very long MCMC.

\begin{table}
\caption{GARCH variance reduction: \small{95\% Confidence interval for the ratio of the variances of ordinary MCMC estimators and ZV-MCMC estimator.}}
\label{table1}
\begin{center}
\begin{tabular}{lccc}
\hline\noalign{\smallskip}
 &$\hat{\omega}_1$ & $\hat{\omega}_2$
& $\hat{\omega}_3$ \\
\noalign{\smallskip}\hline\noalign{\smallskip}
1st Degree $P(x)$  &  8-18 &   13-28 &  12-27 \\
2nd Degree $P(x)$  &  1200-2700 &   6100-13500  &  6200-13800 \\
 3rd Degree $P(x)$  & 21000-47000 &   48000-107000 &   26000-58000  \\
\noalign{\smallskip}\hline
\end{tabular}
\end{center}
\end{table}

Finally, note that the ZV strategy can be used in great generality and can be applied also to more complex GARCH models (such as E-GARCH, I-GARCH, Q-GARCH, GJR-GARCH,  \cite{Bollerslev2010}), provided it is possible to
 analitically compute the necessary derivatives and verify the hypotheses needed
for unbiasedness and CLT, in a way similar to the proof reported in Appendix C.

\section{Discussion}
Cross-fertilizations between physics and statistical literature have proved to be quite effective in the past, especially in the MCMC framework. The first paradigmatic example
is the paper by \cite{Hastings70} first and
\cite{GelfandSmith1990} later on.

Besides translating into statistical terms the paper by \cite{AssarafCaffarel99}, the main effort of our work has been the discussion of unbiasedness and convergence of the ZV-MCMC estimator.
The study of CLT leads to the condition of finiteness for
$\mathbb E_\pi [(\frac{\partial \log\pi(\mathbf x)}{\partial \mathbf x})^2]$.
This quantity has also been used in the recent paper by \cite{Girolami10}
as a metric tensor to improve efficiency in  Langevin diffusion and Hamiltonian
MC methods. Their idea is to choose this metric as an optimal, local tuning of the dynamic,
which is able to take into account the intrinsic anisotropy in the model considered. In our understanding, what makes
these methods and our
extremely efficient, is the common strategy of exploiting information contained in the derivatives
of the log-target. A combination of the two strategies could be explored: once the derivatives of the log-target are computed, they can
be used both to boost the performance of the
Markov chain (as suggested by \cite{Girolami10})
and to achieve variance reduction by using them to design control variates.
This is particularly
easy since control variates can be constructed by simply post-processing the Markov chain and, thus, there is no need to re-run the simulation.

The second main contribution of this paper is the critical discussion of the selection of $H$ and $\psi$.
A  comparison between the variance reduction framework exploited in \cite{Dellaportas2010}
and the choice of different operators $H$ in our context has remarked contras
and benefits of the two approaches.
Different choices of $H$ and $\psi$ could provide alternative efficient variance reduction strategies.
This can be easily achieved by considering a wider class of trial functions: $\psi(\mathbf x)=P(\mathbf x)q(\mathbf  x)$, where,
as before, $P(\mathbf x)$ denotes a parametric class of polynomials, and $q(\mathbf x)$ is an arbitrary (sufficiently regular) function.

In the present research we have explored $\psi$ based on first, second and
third degree polynomials. Despite the use of this fairly restrictive class of trial functions, the degree of variance reduction obtained in the examples in Section \ref{exa} and in other simulation studies (not reported here) is impressive and of the order of ten times (for first degree polynomials) and of thousand times
(for higher degree polynomials), with practically small extra CPU
time needed in the simulation.

Finally, mention should be made to an alternative, more general renormalized function $\tilde f$ reported
in the paper by \cite{AssarafCaffarel2003}, defined as:
\begin{equation}
\tilde f = f +\frac{H\psi}{\sqrt{\pi}}- \frac{\psi(H\sqrt{\pi})}{\pi},\label{tilde2}
\end{equation}
where, again, $H$ is an Hamiltonian operator and $\psi$ a quite arbitrary trial function.
In this setting, if $H=-\frac12\Delta+V$, under the same, mild conditions discussed in Section \ref{sec_unb},
$\tilde f$ has the same expectation as $f$ under $\pi$. This is true without imposing condition (\ref{h}), so that now $V$ can be also chosen arbitrarily. Therefore, the re-normalization (\ref{tilde2}) allows for
a more general class of Hamiltonians.
\section*{Supplementary Materials}
Supplementary materials are available. In Appendices A, B and C the zero variance estimator and the proof of CLT are given for all the examples.
In Appendix D computations of unbiasedness conditions discussed in Section \ref{sec_unb}
are reported and verified for the three examples.

\section*{Acknowledgement}
Thanks are due to D. Bressanini, for bringing to our attention the
paper by Assaraf and Caffarel and helping us translate it into statistical terms; to prof. E. Regazzini and F. Nicola, for discussing the CLT conditions
for the examples; P. Tenconi, F. Carone and F. Leisen for comments and contributions to a preliminary version of this research.
And finally, Assaraf and Caffarel themselves have given us interesting and useful comments that have greatly improved the paper.

\section*{Appendix A: Probit model}
\subsection*{Mathematical formulation}
Let $y_i$ be Bernoulli r.v.'s:
    $y_i|\textbf{x}_i \sim \mathcal{B}(1, p_i), \ \ p_i = \Phi(\textbf{x}_i^T \beta)$,
where $\beta \in \mathbb{R}^d$ is the vector of parameters of the model and $\Phi$ is the c.d.f. of a standard normal
distribution. The likelihood function is:
\begin{eqnarray}
    \l(\beta|\textbf{y}, \textbf{x}) \propto \prod_{i=1}^{n} \left[\Phi(\textbf{x}_i^T \beta)\right]^{y_i} \left[1-\Phi(\textbf{x}_i^T \beta)\right]^{1-y_i}. \nonumber
\end{eqnarray}
As it can be seen by inspection, the likelihood function is invariant under the transformation $(\textbf{x}_i, y_i) \rightarrow (-\textbf{x}_i, 1-y_i)$. Therefore, for the sake of simplicity, in the rest of the example we assume $y_i = 1$ for any $i$, so that the likelihood simplifies:
\begin{eqnarray}
    \l(\beta|\textbf{y}, \textbf{x}) \propto \prod_{i=1}^{n} \Phi(\textbf{x}_i^T \beta).  \nonumber
\end{eqnarray}
This formula shows that the contribution of $\textbf{x}_i = \textbf{0}$ is just a constant $\Phi(\textbf{x}_i^T \beta) = \Phi(0) = \frac{1}{2}$, therefore, without loss of generality, we assume for all $i$, $\textbf{x}_i \neq \textbf{0}$.

Using flat priors, the posterior of the model is proportional to the likelihood,
and the Bayesian estimator of each parameter,  $\beta_k$, is
the expected value of $f_k(\beta) = \beta_k$  under $\pi$ ($k=1, 2, \cdots, d$).

Using Schr\"odinger-type Hamiltonians, $H$ and $\psi_k(\beta) = P_k(\beta) \sqrt{\pi(\beta)}$, as the trial functions, where $P_k(\beta) = \sum_{j=1}^{d} a_{j, k} \beta_j$ is a first degree polynomial, one gets:
\[
    \widetilde{f}_k(\beta) = f_k(\beta) + \frac{H \psi_k(\beta)}{\sqrt{\pi(\beta|\textbf{y}, \textbf{x})}}
                           = f_k(\beta) + \sum_{j=1}^{d} a_{j, k} z_j,\]
where, for $j=1,2,\ldots,d$,
%
        \[
    z_j   =  -\frac{1}{2} \sum_{i=1}^{n} \frac{x_{ij} \phi(\textbf{x}_i^T \beta)}{\Phi(\textbf{x}_i^T \beta)},
\]
because of the assumption $y_i=1$ for any $i$.
\subsection*{Central limit theorem}
In the following, it is supposed that $P$ is a linear polynomial.
In the Probit model, the ZV-MCMC estimators obey a CLT if $z_j$ have finite $2+\delta$ moment under $\pi$, for some $\delta>0$:
\begin{eqnarray}
    \mathbb{E}_{\pi}\left[|z_j|^{2+\delta}\right] &=& c_1 \mathbb{E}_{\pi}\left[ \left| \sum_{i=1}^{n} \frac{x_{ij} \phi(\textbf{x}_i^T \beta)}{\Phi(\textbf{x}_i^T \beta)} \right|^{2+\delta} \right] \nonumber\\
                                       &=& c_1 c_2 \int_{\mathbb{R}^d} \left| \sum_{i=1}^{n} \frac{x_{ij} \phi(\textbf{x}_i^T \beta)}{\Phi(\textbf{x}_i^T \beta)} \right|^{2+\delta} \prod_{i=1}^{n} \Phi(\textbf{x}_i^T \beta) d\beta < \infty.    \nonumber
\end{eqnarray}
where $c_1 = 2^{-2-\delta}$, and $c_2$ is the normalizing constant of $\pi$ (the target posterior).
Define:
\begin{eqnarray}
    K_1(\beta) &=& \left| \sum_{i=1}^{n} \frac{x_{ij} \phi(\textbf{x}_i^T \beta)}{\Phi(\textbf{x}_i^T \beta)} \right|^{2+\delta},     \nonumber\\
    K_2(\beta) &=& \prod_{i=1}^{n} \Phi(\textbf{x}_i^T \beta),     \nonumber\\
    K(\beta)   &=& K_1(\beta) K_2(\beta)     \nonumber
\end{eqnarray}
and therefore:
\begin{eqnarray}
    \mathbb{E}_{\pi}\left[|z_j|^{2+\delta}\right] &=& c \int_{\mathbb{R}^d} K_1(\beta) K_2(\beta) d\beta. \nonumber
\end{eqnarray}
where $c = c_1 c_2$.
Before studying the convergence of this integral, the following property of the likelihood for the probit model is needed.
\begin{proposition}\label{MLE}
Existence and uniqueness of MLE implies that, for any $\beta_0 \in \mathbb{R}^d\setminus\{\mathbf 0\}$, there exists $i$ such that $\textbf{x}^T_i \beta_0 < 0$. \end{proposition}
\noindent
\textbf{Proof} (by contradiction).
Uniqueness of MLE implies that $\textbf{x}^T \textbf{x}$ is full rank, that is, there is no $\beta_0$ orthogonal to all observations $\textbf{x}_i$. This can be seen by contradiction: singularity of $\textbf{x}^T \textbf{x}$ implies existence of a non-zero $\beta_0$ orthogonal to all observations $\textbf{x}_i$. This fact, in turn, implies $l(\beta| \textbf{x}, \textbf{y}) = l(\beta+c\beta_0| \textbf{x}, \textbf{y})$ and, therefore, $l(\bullet| \textbf{x}, \textbf{y})$ does not have a unique global maximum. \\
Next, assume there exists some $\beta_0 \in \mathbb{R}^d$ such that, for any $i$, $\textbf{x}^T_i \beta_0>0$. Then $\beta_0$ is a direction of recession for the negative log-likelihood function $- \sum_{i=1}^{n} \ln \Phi(\textbf{x}^T_i \beta)$ (that is a proper closed convex function). This implies that this function does not have non-empty bounded minimum set (\cite{Rockafellar1970}), which means that the MLE does not exist.$\hfill\blacksquare$\\



\noindent Now, rewriting $\int_{\mathbb R^d}K(\beta)d\beta$ in hyper-spherical
coordinates through the bijective transformation
$(\rho,\theta_1,\ldots, \theta_{d-1}):=F(\mathbf \beta)$, where
$F^{-1}$ is defined as


\begin{equation}\left\{\begin{array}{lcl}\beta_1 &=&\rho\cos(\theta_1)\\[12pt]
\beta_l &=& \rho \cos(\theta_l) \prod_{m=1}^{l-1} \sin(\theta_m), \ \ \ \textmd{ for } l=2, ..., d-1
\\[12pt]
\beta_d &=& \rho \prod_{m=1}^{d-1} \sin(\theta_m), \end{array}\right.\label{spheric}\end{equation}
for $\theta\in\Theta:=\{0\le\theta_i\le\pi$, $i=1,\ldots, d-2$, $0\le\theta_{d-1}<2\pi\}$ and $\rho>0$,
one gets

\[\begin{array}{rcl}\displaystyle\int_{\mathbb R^d}K(\beta)d\beta &=&\displaystyle\int_{\Theta}
\displaystyle\int_0^{+\infty}K(F^{-1}(\rho,\mathbf\theta))
\rho^{d-1}\prod_{j=2}^{d-2}\sin^{d-j}(\theta_{j-1}) \ d \rho d\theta.\\[12pt]
&\le& \displaystyle\int_{{\Theta}}
\displaystyle\int_0^{+\infty}
K(F^{-1}(\rho,\mathbf \theta)) \rho^{d-1}\ d \rho d\theta\\
[12pt]&:=& \displaystyle\int_{\Theta}A(\mathbf\theta)d\theta,\end{array}
\]

%
Observe that the integrand is well defined for any $(\rho,\theta)$ on the
domain of integration, so it is enough to study its asymptotic behaviour when $\rho$ goes to infinity, and
$\mathbf\theta\in\Theta$.

First, analyze
\[K_1(F^{-1}(\rho,\mathbf \theta))=\left| \sum_{i=1}^{n} \frac{x_{ij} \phi(|\textbf{x}_i| \rho\lambda_i(\mathbf \theta)}{\Phi(|\textbf{x}_i|\rho\lambda_i(\mathbf\theta))} \right|^{2+\delta},    \]
where, for any $i$, $\lambda_i$ is a suitable function of the angles $\mathbf \theta$ such that
$\lambda_i\in[-1,1]$, which takes into account the sign of the scalar product in the original coordinates system.

For any $i$, when $\rho\rightarrow\infty$
\begin{itemize}
    \item
    if $\lambda_i<0$, $\frac{x_{ij} \phi(|\textbf{x}_i|\rho\lambda_i)}{\Phi(|\textbf{x}_i|\rho\lambda_i)} \in \mathcal{O} \left(\rho \right)$;
    \item
    if $\lambda_i>0$, $\frac{x_{ij} \phi(|\textbf{x}_i| \rho \lambda_i )}{\Phi(|\textbf{x}_i| \rho)\lambda_i} \in \mathcal{O} \left( \phi(\lambda_i \rho) \right)$;
    \item
    if $\lambda_i= 0$, $\frac{x_{ij} \phi(|\textbf{x}_i| \rho\lambda_i )}{\Phi(|\textbf{x}_i| \rho\lambda_i)} = x_{ij} \sqrt{\frac{2}{\pi}} \in \mathcal{O} \left( 1 \right)$.
\end{itemize}
Therefore:
\begin{eqnarray}
    \sum_{i=1}^{n} \frac{x_{ij} \phi(|\textbf{x}_i| \rho\lambda_i)}{\Phi(|\textbf{x}_i| \rho\lambda_i)} &\in&   \mathcal{O} \left( \rho \right)    \nonumber
\end{eqnarray}
and, for any $\mathbf \theta\in\Theta$:
$    K_1(F^{-1}(\rho,\mathbf\theta)) \in  \mathcal{O} \left( \rho^{2+\delta} \right). $
Now, focus on $K_2(F^{-1}(\rho,\mathbf\theta)) =\prod_{i=1}^{n} \Phi(|\textbf{x}_i| \rho\lambda_i(\mathbf\theta))$;
existence of MLE for the probit model implies that, for any $\mathbf\theta\in\Theta$,
there exists some $l$ ($1 \leq l \leq n$), such that
$\lambda_l(\mathbf\theta)<0$, and therefore:
\begin{equation}
    K_2(F^{-1}(\rho,\mathbf\theta)) < \Phi(|\textbf{x}_l| \rho\lambda_l)
               \in  \mathcal{O} \left( \phi(\lambda_l \rho) \right) \ \  \rho\rightarrow\infty.
               \label{k2}
\end{equation}
Putting these results together leads to
\begin{equation}
    K(F^{-1}(\rho,\mathbf\theta))   = K_1(F^{-1}(\rho,\mathbf\theta)) K_2(F^{-1}(\rho,\mathbf\theta)) \in  \mathcal{O} \left( \rho^{2+\delta} \phi(\lambda_l(\mathbf\theta) \rho) \right) \nonumber
\end{equation}
so that, for any $\mathbf\theta\in\Theta$,
\[K(F^{-1}(\rho,\mathbf \theta))\rho^{d-1}
 \in \mathcal{O}\left( \rho^{1+\delta+d} \phi\left(\lambda_l(\mathbf\theta)\rho\right)\right),\ \ \ \rho\rightarrow +\infty.\]
Therefore, whenever the value $\mathbf\theta\in\Theta$,
its integrand converges to zero rapidly
enough when $\rho\rightarrow +\infty$. This concludes the proof.\\


\textbf{Note 1}. In (\cite{Speckman2009}) it is shown that the existence of the posterior under flat priors for probit and logit models is equivalent to 
the existence and finiteness of MLE. This ensures us that the posterior is well defined in our context. In order to verify the existence of the posterior mean, we can use a simplified version of the proof given above. In other words we should show $\int \beta_j \prod_{i=1}^{n} \Phi(\textbf{x}_i^T \beta) d\beta < +\infty$, that is, $K_1(\beta) = \beta_j \in \mathcal{O}(\rho)$. Therefore, a weaker version of the proof given above can be employed.\\

\textbf{Note 2}. In the proof given above we have used flat priors: although this assumption simplifies the proof, however a very similar proof can be applied for non-flat priors. Assume the prior is $\pi_0(\beta)$. Under this assumption the posterior is
$ \pi(\beta) = \pi_0(\beta) \ \l(\beta|\textbf{y}, \textbf{x}) \propto \pi_0(\beta) \ \prod_{i=1}^{n} \Phi(\textbf{x}_i^T \beta)$
and the control variates are:
$z_j = -\frac{1}{2} \frac{d \ln \pi_0(\beta)}{d \beta_j} -\frac{1}{2} \sum_{i=1}^{n} \frac{x_{ij} \phi(\textbf{x}_i^T \beta)}{\Phi(\textbf{x}_i^T \beta)}$. Therefore we need to prove the $2+\delta$-th moment of $z_j$ under $\pi(\beta)$ is finite. A sufficient condition for this is the finiteness of $2+\delta$-th moments of $-\frac{1}{2} \frac{d \ln \pi_0(\beta)}{d \beta_j}$ and $-\frac{1}{2} \sum_{i=1}^{n} \frac{x_{ij} \phi(\textbf{x}_i^T \beta)}{\Phi(\textbf{x}_i^T \beta)}$ under $\pi(\beta)$. If we assume $\pi_0(\beta)$ is bounded above, the 
latter is a trivial consequence of the proof given above for the flat priors. Therefore we only need to prove the finiteness of the integral
$$
\int_{\mathbb{R}^d} \left| \frac{d \ln \pi_0(\beta)}{d \beta_j} \right|^{2+\delta} \pi_0(\beta) \ \prod_{i=1}^{n} \Phi(\textbf{x}_i^T \beta) d\beta.
$$
Again if we assume the prior is bounded from above, a sufficient condition for the existence of this integral is the existence of the following integral:
$$
\int_{\mathbb{R}^d} \left| \frac{d \ln \pi_0(\beta)}{d \beta_j} \right|^{2+\delta} \ \prod_{i=1}^{n} \Phi(\textbf{x}_i^T \beta) d\beta
$$
A proof very similar to the one given above will show that this integral is finite for common choices of priors $\pi_0(\beta)$ (such as Normal, Student's T, etc).

\section*{Appendix B: Logit model}
\subsection*{Mathematical formulation}
In the same setting as the probit model, let $p_i = \frac{\exp(\textbf{x}_i^T \beta)}{1+\exp(\textbf{x}_i^T \beta)}$
where $\beta \in \mathbb{R}^d$ is the vector of parameters of the model. The likelihood function is:
\begin{eqnarray}
    \l(\beta|\textbf{y}, \textbf{x}) \propto \prod_{i=1}^{n} \left(\frac{\exp(\textbf{x}_i^T \beta)}{1+\exp(\textbf{x}_i^T \beta)}\right)^{y_i}  \left(\frac{1}{1+\exp(\textbf{x}_i^T \beta)}\right)^{1-y_i}. \label{lik_log}
\end{eqnarray}
By inspection, it is easy to verify that the likelihood function is invariant under the transformation:$
    (\textbf{x}_i, y_i) \rightarrow (-\textbf{x}_i, 1-y_i)$.
Therefore, for the sake of simplicity, in the sequel
we assume $y_i = 0$ for any $i$, so that the likelihood simplifies as:
\begin{eqnarray}
    \l(\beta|\textbf{y}, \textbf{x}) \propto \prod_{i=1}^{n} \frac{1}{1+\exp(\textbf{x}_i^T \beta)}.  \nonumber
\end{eqnarray}
The contribution of $\textbf{x}_i = \textbf{0}$ to the likelihood is just a constant, therefore, without loss of generality, it is assumed that $\textbf{x}_i \neq \textbf{0}$ for all $i$.
%
Using flat priors, the posterior distribution is proportional to \eqref{lik_log}
and the Bayesian estimator of each parameter, $\beta_k$, is the expected value of $f_k(\beta) = \beta_k$  under $\pi$ ($k=1, 2, \cdots, d$).
Using the same pair of operator $H$ and test function $\psi_k$ as in Appendinx A, the control variates are:
\begin{eqnarray}
    z_j = \frac{1}{2} \sum_{i=1}^{n} x_{ij} \frac{ \exp(\textbf{x}_i^T \beta) }{1+\exp(\textbf{x}_i^T \beta)},\ \ \  \textmd{for } \ \ \ \ j = 1, 2, \ldots, d. \nonumber
\end{eqnarray}
\subsection*{Central limit theorem}
As for the probit model, the ZV-MCMC estimators obey a CLT if  the
control variates $z_j$ have finite $2 + \delta$ moment under
$\pi$, for some $\delta>0: $\begin{eqnarray}
    \mathbb{E}_{\pi}\left[z_j^{2+\delta}\right] &=& c_1 \mathbb{E}_{\pi}\left[ \left( \sum_{i=1}^{n} x_{ij} \frac{ \exp(\textbf{x}_i^T \beta) }{1+\exp(\textbf{x}_i^T \beta)} \right)^{2+\delta} \right] \nonumber\\
                                       &=& c_1 c_2 \int_{\mathbb{R}^d} \left( \sum_{i=1}^{n} x_{ij} \frac{ \exp(\textbf{x}_i^T \beta) }{1+\exp(\textbf{x}_i^T \beta)} \right)^{2+\delta} \prod_{i=1}^{n} \frac{1}{1+\exp(\textbf{x}_i^T \beta)} d\beta\nonumber
\end{eqnarray}
where $c_1 = 2^{-2-\delta}$, and $c_2$ is the normalizing constant of $\pi$.
The finitiness of the integral is, indeed, trivial. Observe that the function $e^y/(1+e^y)$ is bounded by 1; therefore,
 \[\mathbb{E}_{\pi}\left[z_j^{2+\delta}\right] \le  \left( \sum_{i=1}^{n} x_{ij}\right)^{2+\delta} <\infty. \]

As for probit model, note that it can easily be shown that the posterior means exist under flat priors. Moreover, a very similar proof (see Note 2 of Appendix A) can be used under a normal or Student's T prior distribution.

\section*{Appendix C: GARCH model}
\subsection*{Mathematical formulation}
We assume that the returns are conditionally normal distributed,
  $r(t) | \mathcal{F}_t \sim \mathcal{N}(0, h_t)$, where $h_t$ is a predictable ($\mathcal{F}_{t-1}$ measurable process):
$    h_t = \omega_1 + \omega_3 h_{t-1} + \omega_2 r_{t-1}^2$,
where $\omega_1>0$, $\omega_2 \geq 0$, and $\omega_3 \geq 0$.
Let $\mathbf r=(r_1,\ldots, r_T)$ be the observed time series.
The likelihood function is equal to:
\begin{eqnarray}
   l\left(\omega_1, \omega_2, \omega_3 | \textbf{r} \right)      &\propto&       \left( \prod_{t=1}^{T} h_t \right)^{-\frac{1}{2}}
                                                                          \exp\left( - \frac{1}{2} \sum_{t=1}^{T} \frac{r_t^2}{h_t} \right)       \nonumber
\end{eqnarray}
and using independent truncated Normal priors for the parameters,
the posterior is:
\begin{eqnarray}
   \pi\left(\omega_1, \omega_2, \omega_3 | \textbf{r} \right)                                                             &\propto&      \exp\left[-\frac{1}{2} \left(\frac{\omega_1^2}{\sigma^2(\omega_1)} +  \frac{\omega_2^2}{\sigma^2({\omega_2})}  +  \frac{\omega_3^2}{\sigma^2({\omega_3})} \right)\right] \left( \prod_{t=1}^{T} h_t \right)^{-\frac{1}{2}}  \exp\left( - \frac{1}{2} \sum_{t=1}^{T} \frac{r_t^2}{h_t} \right). \nonumber
\end{eqnarray}
Therefore,
the control variates (when the trial function is a first degree polynomial)
are:
\[
   \frac{\partial \ln \pi}{\partial \omega_i} = - \frac{\omega_i}{\sigma^2({\omega_i})}
                                                - \frac{1}{2}  \sum_{t=1}^{T} \frac{1}{h_t} \ \frac{\partial h_t}{\partial \omega_i}
                                                + \frac{1}{2}  \sum_{t=1}^{T} \frac{r_t^2}{h_t^2}  \ \frac{\partial h_t}{\partial \omega_i}, \ \ \ i=1,2,3, \]

where:
\[    \frac{\partial h_t}{\partial \omega_1} =\frac{1-\omega_3^{t-1}}{1-\omega_3}, \ \  \ \frac{\partial h_t}{\partial \omega_2} =
\left(r_{t-1}^2 + \omega_3 \frac{\partial h_{t-1}}{\partial \omega_2}\right) \mathbb I_{t>1},  \ \ \ \frac{\partial h_t}{\partial \omega_3} =               \left(h_{t-1} + \omega_3 \frac{\partial h_{t-1}}{\partial \omega_3} \right)\mathbb I_{t>1}.\]
\subsection*{Central limit theorem}
In order to prove the CLT for the ZV-MCMC estimator in the Garch
model, we need:
\begin{equation}\dfrac{\partial \ln \pi}{\partial \omega_2},\ \
\dfrac{\partial \ln \pi}{\partial \omega_3}, \ \ \dfrac{\partial \ln
\pi}{\partial \omega_1}\in
L^{2+\delta}(\pi).\label{into}\end{equation}
To this end, $h_t$ and its partial derivatives should be expressed
as a function of $h_0$ and $\mathbf r$:
\[
h_t=\omega_1(\sum_{k=1}^{t-1}1+\omega_3^k)+\omega_3^t h_0 +
\omega_2(\sum_{k=1}^{t-1}1+\omega_3^kr_{t-1-k}^2), \]

\[
\begin{array}{rcl}
\dfrac{\partial \ln h_t}{\partial \omega_1}&=&\frac{1-\omega_3^{t-1}}{1-\omega_3}\mathbb I_{\{t>1\}},\\[12pt]
\dfrac{\partial \ln h_t}{\partial \omega_2} &=& \left(r_{t-1}^2+\displaystyle\sum_{j=0}^{t-2}\omega_3^{t-1-j}r_j^2 \right)\mathbb I_{\{t>1\}},\\[12pt]
\dfrac{\partial \ln h_t}{\partial \omega_3}&=&\left( h_{t-1}+\displaystyle\sum_{j=0}^{t-2}\omega_3^{t-1-j}h_j\right) \mathbb I_{\{t>1\}}.\\[12pt]
\end{array}
\]

 Next, moving
to spherical coordinates, the integral (\ref{into}) can be written
as
\[\int_{[0,\pi/2]^2} \int_0^\infty K_{j}(\rho;\theta,\phi)d\rho d\theta d\phi:=\int_{[0,\pi/2]^2} A_j(\theta,\phi)d\theta d\phi,\]
where, for $j=1,2,3$, $K_j(\cdot;\theta,\phi)=|W_j|^{2+\delta}\times W$, with
\begin{equation}
\begin{array}{rcl} W_1&=&-\frac1{\sigma^2(\omega_1)}\rho\cos\theta\sin\phi-
\dfrac12\displaystyle\sum_{t=2}^T\left(\dfrac1{\tilde h_t}-\dfrac{r_t^2}{\tilde
h_t^2}\right)\frac{1-\rho^{t-1}\cos^{t-1}\phi}{1-\rho\cos\phi},\\[12pt]
W_2&=&-\frac1{\sigma^2(\omega_2)}\rho\sin\theta\sin\phi-
\dfrac12\displaystyle\sum_{t=2}^T\left(\dfrac1{\tilde
h_t}-\dfrac{r_t^2}{\tilde h_t^2}\right)\left(
r_{t-1}^2+\sum_{j=0}^{t-2}x_3^{t-1-j}r_j^2\right),
\\[12pt]
W_3&=&-\frac1{\sigma^2(\omega_3)}\rho\cos\phi-
\dfrac12\displaystyle\sum_{t=2}^T\left(\dfrac1{\tilde h_t}-\dfrac{r_t^2}{\tilde
h_t^2}\right)\left(\tilde h_{t-1}+\sum_{j=0}^{t-2}x_3^{t-1-j}\tilde h_j\right),\\[12pt]
W&=& \exp\left(-\frac12\rho^2
(\frac1{\sigma^2(\omega_1)}\cos^2\theta\sin^2\phi
+\frac1{\sigma^2(\omega_2)}\sin^2\theta\sin^2\phi+\frac1{\sigma^2(\omega_3)}\cos^2\phi
)-\frac12\displaystyle\sum_{t=1}^T\frac{r_t^2}{\tilde
h_t^2}\right)\times\\[12pt]
&&\times \rho^2 \sin\theta\left(\displaystyle\prod_{t=1}^T \tilde
h_t\right)^{-\frac12}
\end{array}\label{k2old}
\end{equation}
and
\[\tilde h_t=-\rho\cos\theta\sin\phi\sum_{k=1}^{t-1}(1+\rho^k\cos^k\phi)+h_0\rho^t\cos^t\phi  +
\rho\sin\theta\sin\phi\sum_{k=1}^{t-1}(1+r_{t-1-k}^2\rho^k\cos^k\phi
).\] The aim is to prove that, for any $\theta, \phi \in[0,\pi/2]$
and for any $j$, $A_j(\theta,\phi)$ is finite. To this end, the
convergence of $A_j$ for any $\theta, \phi$ should be discussed.

Let us study the proper domain of $K_j(\cdot;\theta,\phi)$.
Observe that $K_j(\cdot;\theta,\phi)$ is not defined whenever
$\tilde h_t=0$ and, if $j=1$ and $\phi\neq\pi/2$, also for
$\rho=1/\cos\phi$. However, the discontinuity of $K_3$ at this
point is removable, so that the domain of $K_3$ can be extended by
continuity also at $\rho=1/\cos\phi$.

Since $\tilde h_t=0$ if and only if $\rho=0$, it can be concluded
that, for any $j$ and for any $\theta,\phi\in [0,\pi/2]$, the
proper domain of $K_j(\cdot;\theta, \phi)$ is
$\text{dom} K_j(\cdot;\theta,\phi)=(0+\infty).$
By fixing the value of $\theta$ and $\phi$, let us study the
limits of $K_j$ when $\rho\rightarrow 0$ and
$\rho\rightarrow+\infty$. Observe that, whatever the values of
$\theta$ and $\phi$ are, $W_j$'s are rationale functions of
$\rho$. Therefore, for any $j$, $|W_j|^{2+\delta}$ cannot grow
towards infinity more than polynomially at the boundary of the
domain. On the other hand, $W$ goes to zero with an exponential
rate both when $\rho\rightarrow 0$ and $\rho\rightarrow+\infty$,
for any $\theta$ and $\phi$. This is sufficient to conclude that,
for any $\theta,\phi\in [0, \pi/2]$, the integral $A_j$ is finite
for $j=1,2,3$ and, therefore, condition $(\ref{into})$ holds and
the ZV estimators for the GARCH model obeys a CLT.

\section*{Appendix D: unbiasedness}

In this Appendix, explicit computations are presented, which were
omitted in Section 5. Moreover, it is proved that all
the ZV-MCMC estimators discussed in Section 8 are
unbiased.

Following the same notations as in Section 5, equation (9) follows because
\[\begin{array}{rcl}
\displaystyle\left\langle\dfrac{H\psi}{\sqrt\pi}\right\rangle &:=& \displaystyle\int_\Omega H\psi\sqrt\pi\\[12pt]
&=&\displaystyle\int_\Omega (V\psi\sqrt\pi-\dfrac12\Delta\psi\sqrt\pi)\\[12pt]
&=&\displaystyle\int_\Omega  V\sqrt\pi\psi-
\dfrac12\displaystyle\int_{\partial\Omega}\sqrt\pi\nabla\psi\cdot
\mathbf n d\sigma+ \dfrac12 \int_\Omega
\nabla\sqrt\pi\cdot\nabla\psi\\[12pt]
&=&\displaystyle\int_\Omega  V\sqrt\pi\psi-
\dfrac12\displaystyle\int_{\partial\Omega}\sqrt\pi\nabla\psi\cdot
\mathbf n d\sigma+
\dfrac12\displaystyle\int_{\partial\Omega}\psi\nabla\sqrt\pi\cdot
\mathbf n d\sigma -\dfrac12 \int_\Omega
\psi\Delta\sqrt\pi\\[12pt]
&=&\displaystyle\int_\Omega (H\sqrt\pi)\psi+ \dfrac12
\displaystyle\int_{\partial\Omega}[\psi\nabla\sqrt\pi-\sqrt\pi\nabla\psi]\cdot
\mathbf n d\sigma\\[12pt] &=&\dfrac12
\displaystyle\int_{\partial\Omega}[\psi\nabla\sqrt\pi-\sqrt\pi\nabla\psi]\cdot
\mathbf n d\sigma.\\[12pt] \end{array}\]

Therefore, $\left\langle\dfrac{H\psi}{\sqrt\pi}\right\rangle=0$ if
$\psi\nabla\sqrt\pi=\sqrt\pi\nabla\psi$ on $\partial\Omega$. Now,
let $\psi=P\sqrt\pi$. Then,
\[\nabla\psi=\sqrt\pi\nabla P+\dfrac{P}{2\sqrt\pi}\nabla\pi,\]
so that $\left\langle\dfrac{H\psi}{\sqrt\pi}\right\rangle=0$ if

\[\pi(\mathbf x)\dfrac{\partial P (\mathbf x)}{\partial x_j}=0, \ \ \forall \mathbf x\in\partial\Omega,\ \ j=1,\ldots, d.\]
When $\pi$ has unbounded support, following the previous
computations integrating over the bounded set $B_r$ and taking the limit for $r\rightarrow\infty$, one gets
\begin{equation}\displaystyle\left\langle\dfrac{H\psi}{\sqrt\pi}\right\rangle = \dfrac12\lim_{r\rightarrow+\infty}
\displaystyle\int_{\partial B_r}\pi\nabla P\cdot \mathbf n
d\sigma. \label{multid}\end{equation} Therefore, unbiasedness in
the unbounded case is reached if the limit appearing in the
right-hand side of \eqref{multid} is zero.

Now, the unbiasedness of the ZV-MCMC estimators exploited in
Section 8 is discussed. To this end, condition
\eqref{multid} should be verified. Let $B_\rho$ be a hyper-sphere
of radius $\rho$ and let $\mathbf n:=\frac1{\rho}\beta$ be its normal
versor. Then, for linear $P$, \eqref{multid} equals zero if, for any
$j=1, \ldots, d$,
\begin{equation}\lim_{\rho\rightarrow+\infty} \frac1\rho\int_{B_\rho} \pi(\mathbf \beta)\beta_j dS=0.\label{eq_int}\end{equation}

The Probit model is first considered. By using the
same notations as in Appendix A, the integral in \eqref{eq_int}
is proportional to
\begin{equation} \lim_{\rho\rightarrow+\infty} \frac1\rho\int_{\Theta}K_2(F^{-1}(\rho,\theta))\rho^d d\theta.
\label{int_cond}
\end{equation}
Note that, because of \eqref{k2},
there exist $\rho_0$ and M such that
\[\begin{array}{rcl}
K_2(F^{-1}(\rho,\theta))\rho^d &\le &M \phi(\lambda_{l(\theta)}\rho)\rho^d  \\[12pt]
&\le& M \phi(\lambda_{l(\theta)}\rho_0)\rho_0^d := G(\theta) \ \ \forall \rho\ge\rho_0. \end{array}\]
Since $G(\theta)\in L^1$, by the dominated convergence theorem a sufficient condition to get unbiasedness is
\begin{equation}
\lim_{\rho\rightarrow+\infty}K_2(F^{-1}(\rho,\theta))\rho^{d-1}=0,
\label{int_cond2}\end{equation}
which is true, because of \eqref{k2} for the Probit model.

We now consider the Logit model for which it is easy to prove that Proposition \ref{MLE} holds.
As done for the Probit model, one can write
\[\mathbb{E}_{\pi}\left[z_j^{2+\delta}\right] \propto
\int_{\mathbb R^d} K_1(\beta) K_2(\beta) d\beta,\]
where
\begin{eqnarray}
  K_1(\beta) &=&   \left( \sum_{i=1}^{n} x_{ij} \frac{ \exp(\textbf{x}_i^T \beta) }{1+\exp(\textbf{x}_i^T \beta)} \right)^{2+\delta},   \nonumber\\
    K_2(\beta) &=&   \prod_{i=1}^{n} \frac{1}{1+\exp(\textbf{x}_i^T \beta)},    \nonumber
\end{eqnarray}
By using the
hyper-spherical change of variables in (\ref{spheric}),
we get
\[
\mathbb{E}_{\pi}\left[z_j^{2+\delta}\right]\propto
\int_\Theta
\int_0^\infty K_1(F^{-1}(\rho,\mathbf \theta))
K_2(F^{-1}(\rho,\mathbf \theta))  \rho^{d-1}\ d \rho:\]
and, as for the Probit model, we must verify Equation (\ref{int_cond}).
Now analyze $K_2(F^{-1}(\rho,\mathbf \theta))$; for any $\theta$,
existence of MLE implies the existence of some $l$ ($1 \leq l \leq
n$), such that $\lambda_l(\theta)>0$, and therefore:
\begin{eqnarray}
    K_2(F^{-1}(\rho,\mathbf \theta)) &=& \prod_{i=1}^{n} \frac{1}{1+\exp(|\textbf{x}_i|\rho \lambda_i)}     \nonumber\\
               &<& \frac{1}{1+\exp(|\textbf{x}_l|\rho\lambda_l)}     \nonumber\\
               &\in&  \mathcal{O} \left( \exp(-\rho \lambda_l) \right).
               \label{k2logit}
\end{eqnarray}

Therefore, there exist $\rho_0$, M such that
\[\begin{array}{rcl}
K_2(F^{-1}(\rho,\theta))\rho^d &\le &M \exp(-\lambda_{l(\theta)}\rho)\rho^d  \\[12pt]
&\le& M \exp(-\lambda_{l(\theta)}\rho_0)\rho_0^d := G(\theta) \ \ \forall \rho\ge\rho_0, \end{array}\]
where $G(\theta)\in L^1$.
These computations allow us to use Equation \eqref{int_cond2} as a sufficient condition to get unbiasedness,
and its proof becomes trivial.

Finally consider the GARCH model. In this
case, $B_\rho$ is the portion of a sphere of radius $\rho$ defined
on the positive orthant. Then, the limit
\[ \lim_{\rho\rightarrow+\infty} \frac1\rho\int_{[0,\pi/2]^{2}}W(F^{-1}(\rho,\theta))\rho^d d\theta, \]
where $W$ was defined in \eqref{k2old}, should be discussed.
Again, an application of the dominated convergence theorem leads to the simpler condition
\[\lim_{\rho\rightarrow+\infty}W(F^{-1}(\rho,\theta))\rho^{2}=0,\]
which is true, since W decays
with an exponential rate.

\bibliographystyle{spmpsci}
\bibliography{ZeroVarianceMCMC2}

\begin{thebibliography}{10}
\providecommand{\url}[1]{{#1}}
\providecommand{\urlprefix}{URL }
\expandafter\ifx\csname urlstyle\endcsname\relax
  \providecommand{\doi}[1]{DOI~\discretionary{}{}{}#1}\else
  \providecommand{\doi}{DOI~\discretionary{}{}{}\begingroup
  \urlstyle{rm}\Url}\fi

\bibitem{Adler1981}
Adler, S.: Over-relaxation method for the {M}onte {C}arlo evaluation of the
  partition function for multiquadratic actions.
\newblock Phys. Rev. D \textbf{23}, 2901--2904 (1981)

\bibitem{AlbertChib03}
Albert, J., Chib, S.: Bayesian analysis of binary and polychotomous response
  data.
\newblock Journal of the American Statistical Association \textbf{88, 422},
  669--679 (1993)

\bibitem{Ardia2008}
Ardia, D.: Financial risk management with bayesian estimation of {GARCH}
  models: Theory and applications.
\newblock In: Lecture Notes in Economics and Mathematical Systems 612.
  Springer-Verlag (2008)

\bibitem{AssarafCaffarel99}
Assaraf, R., Caffarel, M.: Zero-{V}ariance principle for {M}onte {C}arlo
  algorithms.
\newblock Physical Review letters \textbf{83, 23}, 4682--4685 (1999)

\bibitem{AssarafCaffarel2003}
Assaraf, R., Caffarel, M.: Zero-variance zero-bias principle for observables in
  quantum {M}onte {C}arlo: Application to forces.
\newblock The Journal of Chemical Physics \textbf{119, 20}, 10,536--10,552
  (2003)

\bibitem{BaroneFrigessi89}
Barone, P., Frigessi, A.: Improving stochastic relaxation for {G}aussian random
  fields.
\newblock Probability in the Engineering and Informational Sciences \textbf{4},
  369--389 (1989)

\bibitem{Baronealii2001}
Barone, P., Sebastiani, G., Stander, J.: General over-relaxation {M}arkov chain
  {M}onte {C}arlo algorithms for {G}aussian densities.
\newblock Statistics \& Probability Letters \textbf{52,2}, 115--124 (2001)

\bibitem{Bollerslev86}
Bollerslev, T.: Generalized autoregressive conditional heteroskedasticity.
\newblock Journal of Econometrics \textbf{31, 3}, 307--327 (1986)

\bibitem{Bollerslev2010}
Bollerslev, T.: Glossary to {A}{R}{C}{H} ({G}{A}{R}{C}{H}).
\newblock In: Volatility and Time Series Econometrics, Essays in Honor of
  Robert Engle, Edited by Tim Bollerslev, Jeffrey Russell and Mark Watson.
  Oxford University Press, Oxford, UK (2010)

\bibitem{Breweralii1996}
Brewer, M., Aitken, C., Talbot, M.: A comparison of hybrid strategies for
  {G}ibbs sampling in mixed graphical models.
\newblock Computational Statistics \textbf{21}, 343--365 (1996)

\bibitem{BrooksGelman1998}
Brooks, S., Gelman, A.: Some issues in monitoring convergence of iterative
  simulations.
\newblock Computing Science and Statistics  (1998)

\bibitem{CraiuLemieux07}
Craiu, R., Lemeieux, C.: Acceleration of the multiple-try {M}etropolis
  algorithm using antithetic and stratified sampling.
\newblock Journal Statistics and Computing \textbf{17, 2}, 109--120 (2007)

\bibitem{CraiuMeng05}
Craiu, R., Meng, X.: Multiprocess parallel antithetic coupling for backward and
  forward {M}arkov chain {M}onte {C}arlo.
\newblock The Annals of Statistics \textbf{33, 2}, 661--697 (2005)

\bibitem{Dellaportas2010}
Dellaportas, P., Kontoyiannis, I.: Control variates for estimation based on
  reversible {M}arkov chain {M}onte {C}arlo samplers.
\newblock Journal of the Royal Statistical Society, Series B. \textbf{74(1)},
  133--161 (2012)

\bibitem{Diaconisalii2000}
Diaconis, P., Holmes, S., Neal, R.F.: Analysis of a nonreversible {M}arkov
  chain sampler.
\newblock Ann. Appl. Probab. \textbf{10,3}, 726--752 (2000)

\bibitem{Duanealii1987}
Duane, S., Kennedy, A., Pendleton, B., Roweth, D.: Hybrid {M}onte {C}arlo.
\newblock Physics Letters B \textbf{195}, 216--222 (2010)

\bibitem{Flury88}
Flury, B., Riedwyl, H.: Multivariate Statistics.
\newblock Chapman and Hall (1988)

\bibitem{Fortalii2003}
Fort, G., Moulines, E., Roberts, G., Rosenthal, S.: On the geometric ergodicity
  of hybrid samplers.
\newblock Journal of Applied Probability \textbf{40, 1}, 123--146 (2003)

\bibitem{GelfandSmith1990}
Gelfand, A., Smith, A.: Sampling-based approaches to calculating marginal
  densities.
\newblock J. American Statistical Association \textbf{85}, 398--409 (1990)

\bibitem{Girolami10}
Girolami, M., Calderhead, B.: Riemannian manifold {L}angevin and {H}amiltonian
  {M}onte {C}arlo methods.
\newblock J. R. Statist. Soc. B \textbf{73, 2}, 1--37 (2011)

\bibitem{GreenHan92}
Green, P., Han, X.: Metropolis methods, {G}aussian proposals, and antithetic
  variables.
\newblock In: P.~Barone, A.~Frigessi, M.~Piccioni (eds.) Lecture Notes in
  Statistics, Stochastic Methods and Algorithms in Image Analysis, vol.~74, pp.
  142--164. Springer Verlag (1992)

\bibitem{GreenMiraRJ01}
Green, P.J., Mira, A.: Delayed rejection in reversible jump
  {M}etropolis-{H}astings.
\newblock Biometrika \textbf{88}, 1035--1053 (2001)

\bibitem{Hastings70}
Hastings, W.K.: Monte {C}arlo sampling methods using {M}arkov chains and their
  applications.
\newblock Biometrika \textbf{57}, 97--109 (1970)

\bibitem{Henderson97}
Henderson, S.: Variance reduction via an approximating {M}arkov process.
\newblock Ph.D. thesis, Department of Operations Research, Stanford University,
  Stanford, CA (1997)

\bibitem{HendersonGlynn2002}
Henderson, S., Glynn, P.: Approximating martingales for variance reduction in
  {M}arkov process simulation.
\newblock Math. Oper. Res. \textbf{27, 2}, 253--271 (2002)

\bibitem{Higdon1998}
Higdon, D.: Auxiliary variable methods for {M}arkov chain {M}onte {C}arlo with
  applications.
\newblock Journal of the American Statistical Association \textbf{93}, 585--595
  (1998)

\bibitem{Ishwaran1999}
Ishwaran, H.: Applications of hybrid {M}onte {C}arlo to {B}ayesian generalized
  linear models: quasicomplete separation and neural networks.
\newblock J. Comp. Graph. Statist. \textbf{8}, 779--799 (1999)

\bibitem{Leisen2010}
Leisen, F., Dalla~Valle, L.: A new multinomial model and a zero variance
  estimation.
\newblock Communications in Statistics - Simulation and Computation
  \textbf{39(4)}, 846--859 (2010)

\bibitem{Linnik59}
Linnik, Y.V.: An information-theoretic proof of the central limit theorem with
  {L}indeberg conditions.
\newblock Theory of Probability and its Applications \textbf{4}, 288--299
  (1959)

\bibitem{Loh1994}
Loh, W.: Methods of control variates for discrete event simulation.
\newblock Ph.D. thesis, Department of Operations Research, Stanford University,
  Stanford, CA (1994)

\bibitem{BayesianCore}
Marin, J.M., Robert, C.: Bayesian Core: A Practical Approach to Computational
  Bayesian Statistics.
\newblock Springer (2007)

\bibitem{MiraGeyer00}
Mira, A., Geyer, C.J.: On reversible {M}arkov chains.
\newblock Fields Inst. Communic.: {M}onte {C}arlo Methods \textbf{26}, 93--108
  (2000)

\bibitem{Miraalii2001}
Mira, A., M{\"o}ller, J., Roberts, G.O.: Perfect slice samplers.
\newblock Journal of the Royal Statistical Soc. Ser. B \textbf{63, 3}, 593--606
  (2001)

\bibitem{TierneyMira2002}
Mira, A., Tierney, L.: Efficiency and convergence properties of slice samplers.
\newblock Scandinavian Journal of Statistics \textbf{29}, 1--12 (2002)

\bibitem{Neal1994}
Neal, R.: An improved acceptance procedure for the hybrid {M}onte {C}arlo
  algorithm.
\newblock Journal of Computational Physics \textbf{111}, 194--203 (1994)

\bibitem{Neal1995}
Neal, R.M.: Suppressing random walks in {M}arkov chain {M}onte {C}arlo using
  ordered overrelaxation.
\newblock Tech. rep., Learning in Graphical Models (1995)

\bibitem{Nelson1989}
Nelson, B.: Batch size effects on the efficiency of control variates in
  simulation.
\newblock European Journal of Operational Research \textbf{2(27)}, 184--196
  (1989)

\bibitem{PhilippeRobert01}
Philippe, A., Robert, C.: Riemann sums for {MCMC} estimation and convergence
  monitoring.
\newblock Statistics and Computing \textbf{11}, 103--105 (2001)

\bibitem{RipleySimulation87}
Ripley, B.: Stochastic Simulation.
\newblock John Wiley \& Sons (1987)

\bibitem{Rockafellar1970}
Rockafellar, R.: Convex analysis, pp. 264--265.
\newblock Princeton University Press (1970)

\bibitem{So06}
So, M.K.P.: Bayesian analysis of nonlinear and non-{G}aussian state space
  models via multiple-try sampling methods.
\newblock Statistics and Computing \textbf{16}, 125--141 (2006)

\bibitem{Speckman2009}
Speckman P.L.~Lee, J., Sun, D.: Existence of the mle and propriety of
  posteriors for a general multinomial choice model.
\newblock Statistica Sinica \textbf{19}, 731--748 (2009)

\bibitem{SwendsenWang1987}
Swendsen, R., Wang, J.: Non universal critical dynamics in {M}onte {C}arlo
  simulations.
\newblock Phys. Rev. Lett. \textbf{58}, 86--88 (1987)

\bibitem{TierneyMCMC94}
Tierney, L.: Markov chains for exploring posterior distributions.
\newblock Annals of Statistics \textbf{22}, 1701--1762 (1994)

\bibitem{TierneyMiraAdaptive99}
Tierney, L., Mira, A.: Some adaptive {M}onte {C}arlo methods for {B}ayesian
  inference.
\newblock Statistics in Medicine \textbf{18}, 2507--2515 (1999)

\bibitem{Vandykalii2001}
Van~Dyk, D., Meng, X.: The art of data augmentation.
\newblock Journal of Computational and Graphical Statistics \textbf{10}, 1--50
  (2001)

\end{thebibliography}


\end{document}